\documentclass[titlepage,12pt]{article}
\usepackage[dvips]{graphicx}

\begin{document}

\title{Effects of a mixed vector-scalar kink-like potential for spinless
particles in two-dimensional spacetime}
\date{}
\author{Antonio S. de Castro\thanks{%
E-mail address: castro@feg.unesp.br (A.S. de Castro)} \\
\\
Universidade de Coimbra\\
Departamento de F\'{\i}sica and Centro de F\'{\i}sica Computacional\\
P-3004-516 Coimbra - Portugal\\
and\\
UNESP - Campus de Guaratinguet\'{a}\\
Departamento de F\'{\i}sica e Qu\'{\i}mica\\
12516-410 Guaratinguet\'{a} SP - Brasil\\
\\
}
\date{}
\maketitle

\begin{abstract}
The intrinsically relativistic problem of spinless particles subject to a
general mixing of vector and scalar kink-like potentials ($\sim \mathrm{tanh}%
\,\gamma x$) is investigated. The problem is mapped into the exactly
solvable Surm-Liouville problem with the Rosen-Morse potential and exact
bounded solutions for particles and antiparticles are found. The behaviour
of the spectrum is discussed in some detail. An apparent paradox concerning
the uncertainty principle is solved by recurring to the concept of effective
Compton wavelength.
\end{abstract}

\section{Introduction}

There has been a continuos interest for solving the Klein-Gordon (KG)
equation in the four-dimensional space-time as well as in lower dimensions
for a variety of potentials \cite{ada}-\cite{cas3}. A few recent works has
been devoted to the investigation of the solutions of the KG equation by
assuming that the vector potential is equal to the scalar potential \cite%
{chao1}-\cite{zha} whereas other works take a more general mixing \cite{ada}-%
\cite{che},\cite{cas1}-\cite{cas3}. Although the KG equation can give
relativistic corrections to the nonrelativistic quantum mechanics, in some
circumstances it can present solutions not found in a nonrelativistic
scheme. Undoubtedly such circumstances reveal to be a powerful tool to
obtain a deeper insight about the nature of the KG equation and its
solutions.

The Dirac equation in the background of the kink configuration of the $\phi
^{4}$ model ($\mathrm{tanh}\,\gamma x$) \cite{raj} is of interest in quantum
field theory where topological classical backgrounds are responsible for
inducing a fractional fermion number on the vacuum. Models of these kinds,
known as kink models are obtained in quantum field theory as the continuum
limit of linear polymer models \cite{gol}-\cite{sem}. In a recent paper the
complete set of bound states of fermions in the presence of this sort of
background has been addressed by considering a pseudoscalar coupling \cite%
{asc}. A peculiar feature of the kink-like potential is the absence of
bounded solutions in a nonrelativistic theory because it gives rise to an
ubiquitous repulsive potential. The spectrum of this problem was found
analytically due to the fact that, apart from solutions corresponding to $%
|E|=mc^{2}$, the problem is reducible to the finite set of solutions of the
nonrelativistic modified P\"{o}schl-Teller potential.

In the present work the problem of a spinless particle in the background of
a kink-like potential is considered with a general mixing of vector and
scalar Lorentz structures. Apart from the intrinsic interest as new
solutions of a fundamental equation in physics, the bound-state solutions of
this system is related to a number of applications ranging from
ferroelectric domain walls in solids, magnetic chains and Josephson
junctions \cite{bra}. The problem is mapped into an exactly solvable
Sturm-Liouville problem of a Schr\"{o}dinger-like equation with an effective
Rosen-Morse potential \cite{ros}-\cite{ach}. The whole spectrum of this
relativistic problem is found analytically, if the particle is massless or
not. Nevertheless, bounded solutions do exist only if the scalar coupling is
stronger than the vector coupling. A remarkable feature of this problem is
the possibility of trapping a particle with an uncertainty in the position
that can shrink to zero for arbitrarily large values of the potential
parameters. Due to the repulsive nature of the kink-like potential those
solutions do not manifest in a nonrelativistic approach even though one can
find $|E|\simeq mc^{2}$. Therefore, they are intrinsically relativistic
solutions of the Klein-Gordon equation.

In the presence of vector and scalar potentials the 1+1 dimensional
time-independent KG equation for a particle of rest mass $m$ reads

\begin{equation}
-\hbar ^{2}c^{2}\,\frac{d^{2}\phi }{dx^{2}}+\left( mc^{2}+V_{s}\right)
^{2}\phi =\left( E-V_{v}\right) ^{2}\phi  \label{1}
\end{equation}

\noindent where $E$ is the energy of the particle, $c$ is the velocity of
light and $\hbar $ is the Planck constant. The vector and scalar potentials
are given by $V_{v}$ and $V_{s}$, respectively. The subscripts for the terms
of potential denote their properties under a Lorentz transformation: $v$ for
the time component of the 2-vector potential and $s$ for the scalar term.
Note that $\phi $ remains invariant under the simultaneous transformations $%
E\rightarrow -E$ and $V_{v}\rightarrow -V_{v}$. Furthermore, for $V_{v}=0$,
the case of a pure scalar potential, the negative- and positive-energy
levels are disposed symmetrically about $E=0$. It is remarkable that the KG
equation with a scalar potential, or a vector potential contaminated with
some scalar coupling, is not invariant under the simultaneous changes $%
V\rightarrow V+const.$ and $E\rightarrow E+const.$, this is so because only
the vector potential couples to the positive-energies in the same way it
couples to the negative-ones, whereas the scalar potential couples to the
mass of the particle. Therefore, if there is any scalar coupling the the
energy itself has physical significance and not just the energy difference.

The KG equation can also be written as
\begin{equation}
-\frac{\hbar ^{2}}{2}\,\phi ^{\prime \prime }+\left( \frac{%
V_{s}^{2}-V_{v}^{2}}{2c^{2}}+mV_{s}+\frac{E}{c^{2}}\,V_{v}\right) \,\phi =%
\frac{E^{2}-m^{2}c^{4}}{2c^{2}}\,\phi  \label{2}
\end{equation}

\noindent

\noindent From this one can see that for potentials which tend to $\pm
\infty $ as $|x|\rightarrow \infty $ it follows that the effective potential
tends to $\left( V_{s}^{2}-V_{v}^{2}\right) /\left( 2mc^{2}\right) $, so
that the KG equation furnishes a purely discrete (continuum) spectrum for $%
|V_{s}|>|V_{v}|$ ($|V_{s}|<|V_{v}|$). On the other hand, if the potentials
remains finite as $|x|\rightarrow \infty $ the continuum spectrum is
omnipresent but the necessary conditions for the existence of a discrete
spectrum is not an easy task for general functional forms. \noindent In the
nonrelativistic approximation (potential energies small compared to $mc^{2}$
and $E\simeq mc^{2}$) Eq. (\ref{2}) becomes

\begin{equation}
\left( -\frac{\hbar ^{2}}{2m}\frac{d^{2}}{dx^{2}}+V_{v}+V_{s}\right) \phi
=\left( E-mc^{2}\right) \phi  \label{3}
\end{equation}

\noindent so that $\phi $ obeys the Schr\"{o}dinger equation with binding
energy equal to $E-mc^{2}$ without distinguishing the contributions of
vector and scalar potentials.

It is well known that a confining potential in the nonrelativistic approach
is not confining in the relativistic approach when it is considered as a
Lorentz vector. It is surprising that relativistic confining potentials may
result in nonconfinement in the nonrelativistic approach. This last
phenomenon is a consequence of the fact that vector and scalar potentials
couple differently in the KG equation whereas there is no such distinction
among them in the Schr\"{o}dinger equation. This observation permit us to
conclude that even a \textquotedblleft repulsive\textquotedblright\
potential can be a confining potential. For both cases $V_{v}=V_{s}$ and $%
V_{v}=-V_{s}$, the KG equation reduces to Schr\"{o}dinger-like equations
with effective potentials with the same functional dependence as $V_{s}$.
The case $V_{v}=-V_{s}$ can present bounded solutions in the relativistic
approach, although it reduces to the free-particle problem in the
nonrelativistic limit. The attractive vector potential for a particle is, of
course, repulsive for its corresponding antiparticle, and vice versa.
However, the attractive (repulsive) scalar potential for particles is also
attractive (repulsive) for antiparticles. For $V_{v}=V_{s}$ and an
attractive vector potential for particles, the scalar potential is
counterbalanced by the vector potential for antiparticles as long as the
scalar potential is attractive and the vector potential is repulsive. As a
consequence there is no bounded solution for antiparticles. For $%
V_{v}=-V_{s} $ and a repulsive vector potential for particles, the scalar
and the vector potentials are attractive for antiparticles but their effects
are counterbalanced for particles. Thus, recurring to this simple standpoint
one can anticipate in the mind that there is no bound-state solution for
particles in this last case of mixing.

\section{The mixed vector-scalar kink-like potential}

Now let us focus our attention on scalar and vector potentials in the form
\begin{equation}
V_{v}=\hbar c\gamma g_{v}\,\mathrm{tanh}\,\gamma x,\quad V_{s}=\hbar c\gamma
g_{s}\,\mathrm{tanh}\,\gamma x  \label{4}
\end{equation}%
\noindent where the skew parameter, $\gamma $, and the dimensionless
coupling constants, $g_{v}$ and $g_{s}$, are real numbers. The potentials
are invariant under the change $\gamma \rightarrow -\gamma $ so that the
results can depend only on $|\gamma |$. In this case the Sturm-Liouville
problem corresponding to Eq. (\ref{2}) becomes

\begin{equation}
H_{\mathtt{eff}}\,\phi =-\frac{\hbar ^{2}}{2m_{\mathtt{eff}}}\,\phi ^{\prime
\prime }+V_{\mathtt{eff}}\,\phi =E_{\mathtt{eff}}\,\phi  \label{5}
\end{equation}

\noindent where one can recognize the effective potential as the exactly
solvable Rosen-Morse potential \cite{ros}-\cite{ach}

\begin{equation}
V_{\mathtt{eff}}=-V_{1}\,\mathrm{sech}^{2}\gamma x+V_{2}\,\mathrm{tanh}%
\,\gamma x  \label{6}
\end{equation}

\begin{equation}
V_{1}=\frac{\hbar ^{2}\gamma ^{2}}{2m_{\mathtt{eff}}}\,\left(
g_{s}^{2}-g_{v}^{2}\right) ,\quad V_{2}=\frac{\hbar \gamma }{m_{\mathtt{eff}%
}c}\left( mc^{2}g_{s}+Eg_{v}\right) \quad  \label{7}
\end{equation}

\begin{equation}  \label{8}
\end{equation}

\noindent The Rosen-Morse potential is a binding potential only if $V_{1}>0$
and $|V_{2}|<2V_{1}$ because only in this circumstance it presents a well
structure with the possible discrete effective eigenenergies into the range
\begin{equation}
-\left( V_{1}+\frac{V_{2}^{2}}{4V_{1}}\right) <E_{\mathtt{eff}}<-|V_{2}|
\label{8a}
\end{equation}%
whereas $E_{\mathtt{eff}}>-|V_{2}|$ corresponds to the continuous part of
the spectrum. Thus, in view of the interest in the bound-state solutions it
is convenient to rewrite the coupling constants in terms of the variable $%
\xi $ defined into the interval $\left( -1,1\right) $ as
\begin{equation}
g_{s}=g,\quad g_{v}=g\sin \left( \frac{\pi \xi }{2}\right)  \label{9}
\end{equation}

\noindent in such a way that
\begin{equation}
V_{1}=\frac{\hbar ^{2}\gamma ^{2}g^{2}}{2m_{\mathtt{eff}}}\cos ^{2}\left(
\frac{\pi \xi }{2}\right) ,\quad V_{2}=\frac{\hbar \gamma g}{m_{\mathtt{eff}%
}c}\left[ mc^{2}+E\sin \left( \frac{\pi \xi }{2}\right) \right]  \label{10}
\end{equation}

\noindent Normalizable KG wave functions, corresponding to bound-state
solutions, are subject to the boundary condition $\phi (\pm \infty)=0$ and
the effective eigenenergy is given by (see Ref. \cite{nie})

\[
E_{\mathtt{eff}}=\frac{E^{2}-m_{\mathtt{eff}}^{2}c^{4}}{2m_{\mathtt{eff}%
}c^{2}}=-\left[ \frac{\hbar ^{2}\gamma ^{2}a_{n} ^{2}}{2m_{\mathtt{eff}}}+%
\frac{m_{\mathtt{eff}}V_{2}^{2}}{2\hbar ^{2}\gamma ^{2}a_{n} ^{2}}\right]
\qquad \qquad \qquad
\]

\begin{equation}
\qquad \qquad \qquad =-\,\frac{1}{2m_{\mathtt{eff}}c^{2}}\left\{ \hbar
^{2}c^{2}\gamma ^{2}a_{n} ^{2}+\frac{g^{2}\left[ mc^{2}+E\sin \left( \frac{%
\pi \xi }{2}\right) \right] ^{2}}{a_{n} ^{2}}\right\}  \label{11}
\end{equation}

\noindent with the effective mass, $m_{\mathtt{eff}}$, defined as
\begin{equation}
m_{\mathtt{eff}}=\sqrt{m^{2}+\left[ \frac{\hbar \gamma g}{c}\cos \left(
\frac{\pi \xi }{2}\right) \right] ^{2}}  \label{14}
\end{equation}%
\noindent In addittion, $a_{n}=s-n$ and

\begin{equation}
s=\frac{1}{2}\left( -1+\sqrt{1+\frac{8m_{\mathtt{eff}}V_{1}}{\hbar
^{2}\gamma ^{2}}}\,\right) =\frac{1}{2}\left( -1+\sqrt{1+\left[ 2g\cos
\left( \frac{\pi \xi }{2}\right) \right] ^{2}}\,\right)  \label{12}
\end{equation}%
\noindent The quantum number $n$ satisfies the constraint equation%
\begin{equation}
n=0,1,2,\ldots \leq s-\sqrt{\frac{m_{\mathtt{eff}}|V_{2}|}{\hbar ^{2}\gamma
^{2}}}=s-\sqrt{\frac{|g\left[ mc^{2}+E\sin \left( \frac{\pi \xi }{2}\right) %
\right] |}{\hbar c|\gamma |}}  \label{13}
\end{equation}

\noindent From Eqs. (\ref{11}) and (\ref{14}) one obtains the second-degree
algebraic equation for the KG energies:

\[
\left[ a_{n}^{2}+g^{2}\sin ^{2}\left( \frac{\pi \xi }{2}\right) \right]
E^{2}+2g^{2}mc^{2}\sin \left( \frac{\pi \xi }{2}\right) E\qquad \qquad
\qquad
\]%
\begin{equation}
\qquad +m^{2}c^{4}\left( g^{2}-a_{n}^{2}\right) +\hbar ^{2}c^{2}\gamma
^{2}a_{n}^{2}\left[ a_{n}^{2}-g^{2}\cos ^{2}\left( \frac{\pi \xi }{2}\right) %
\right] =0  \label{15}
\end{equation}%
\noindent whose solutions are
\[
E=-\frac{mc^{2}g^{2}\sin \left( \frac{\pi \xi }{2}\right) }{%
a_{n}^{2}+g^{2}\sin ^{2}\left( \frac{\pi \xi }{2}\right) }\qquad \qquad
\qquad \qquad \qquad \qquad \qquad \qquad \qquad \qquad \qquad \qquad \qquad
\]

\begin{equation}
\pm \frac{a_{n}\,c\sqrt{m^{2}c^{2}\left[ a_{n}^{2}-g^{2}\cos ^{2}\left(
\frac{\pi \xi }{2}\right) \right] +\hbar ^{2}\gamma ^{2}\left[ \frac{g^{4}}{4%
}\sin ^{2}\left( \pi \xi \right) +a_{n}^{2}\,g^{2}\cos \left( \pi \xi
\right) -a_{n}^{4}\right] }}{a_{n}^{2}+g^{2}\sin ^{2}\left( \frac{\pi \xi }{2%
}\right) }  \label{16}
\end{equation}

\begin{figure}[th]
\begin{center}
\includegraphics[width=10cm, angle=270]{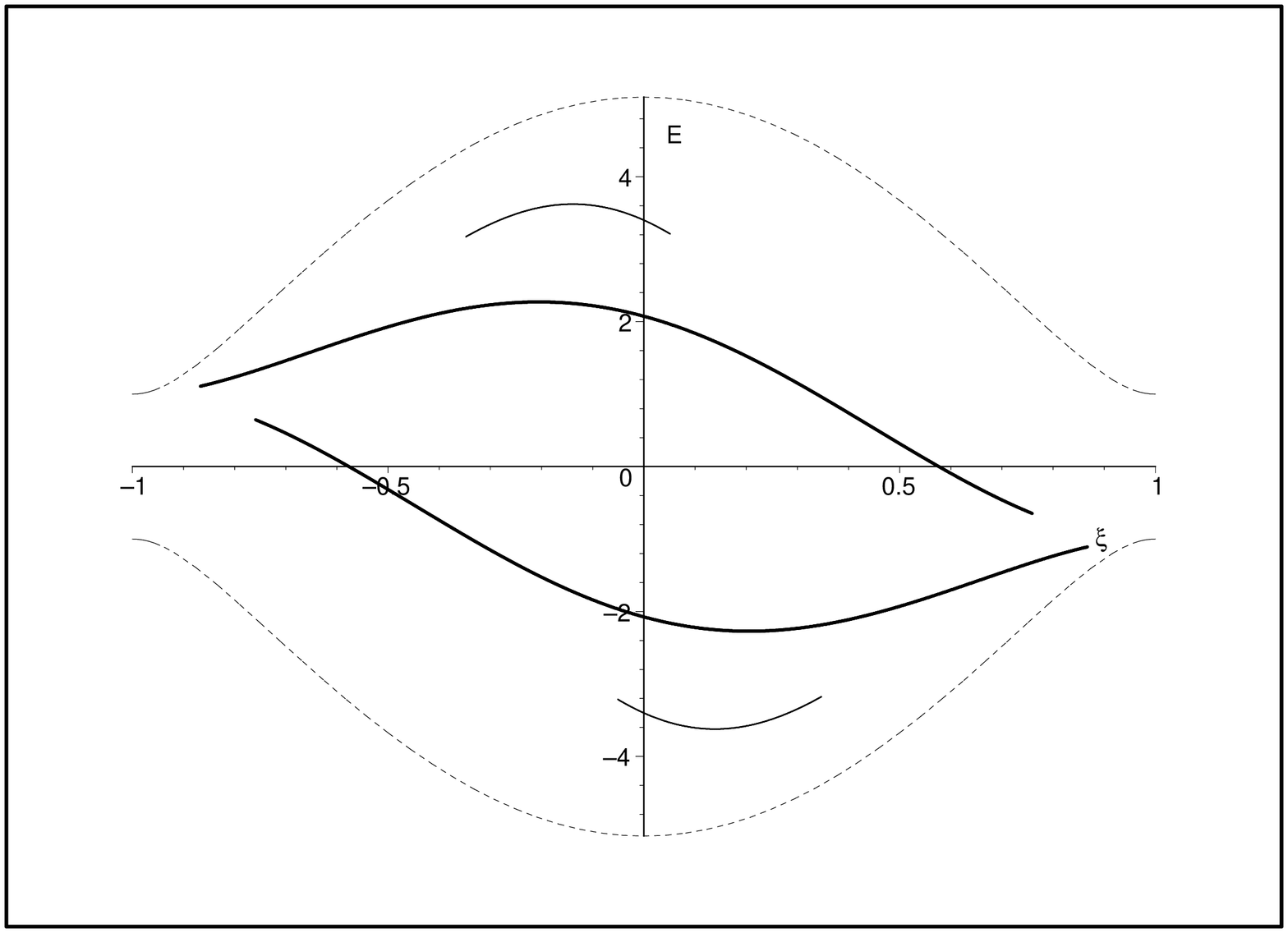}
\end{center}
\par
\vspace*{-0.1cm}
\caption{The KG energies for the kink-like potential as a function of $%
\protect\xi $. The full thick line stands for for $n=0$, the full thin line
for $n=1$ and the dashed line for $\pm m_{\mathtt{eff}}$ ($\hbar =c=m=%
\protect\gamma =1$ and $g=5$). }
\label{Fig1}
\end{figure}

\begin{figure}[th]
\begin{center}
\includegraphics[width=10cm, angle=270]{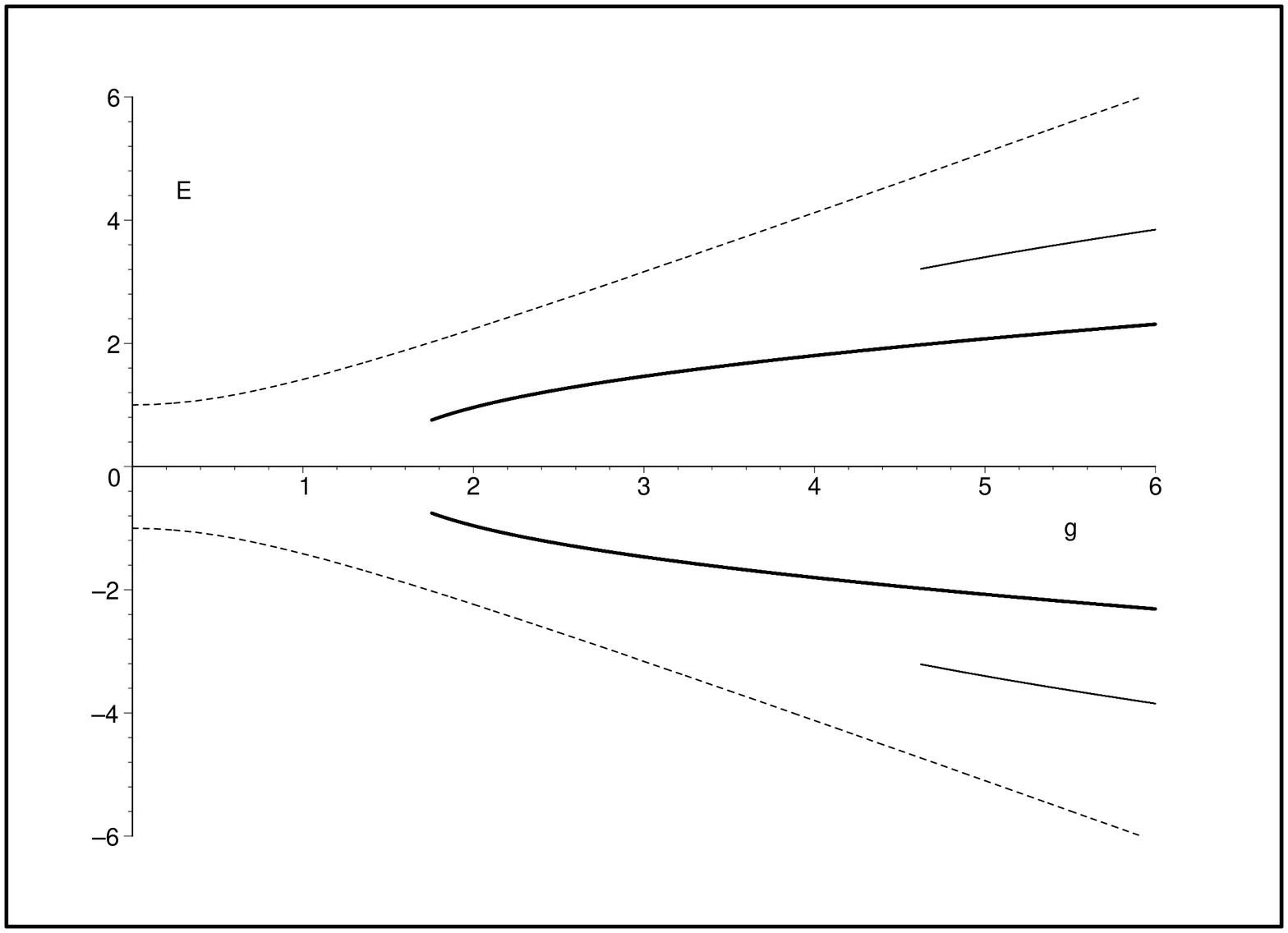}
\end{center}
\par
\vspace*{-0.1cm}
\caption{The same as in Fig. 1 as a function of $g$ for $\protect\xi =0$.}
\label{Fig2}
\end{figure}

\begin{figure}[th]
\begin{center}
\includegraphics[width=10cm, angle=270]{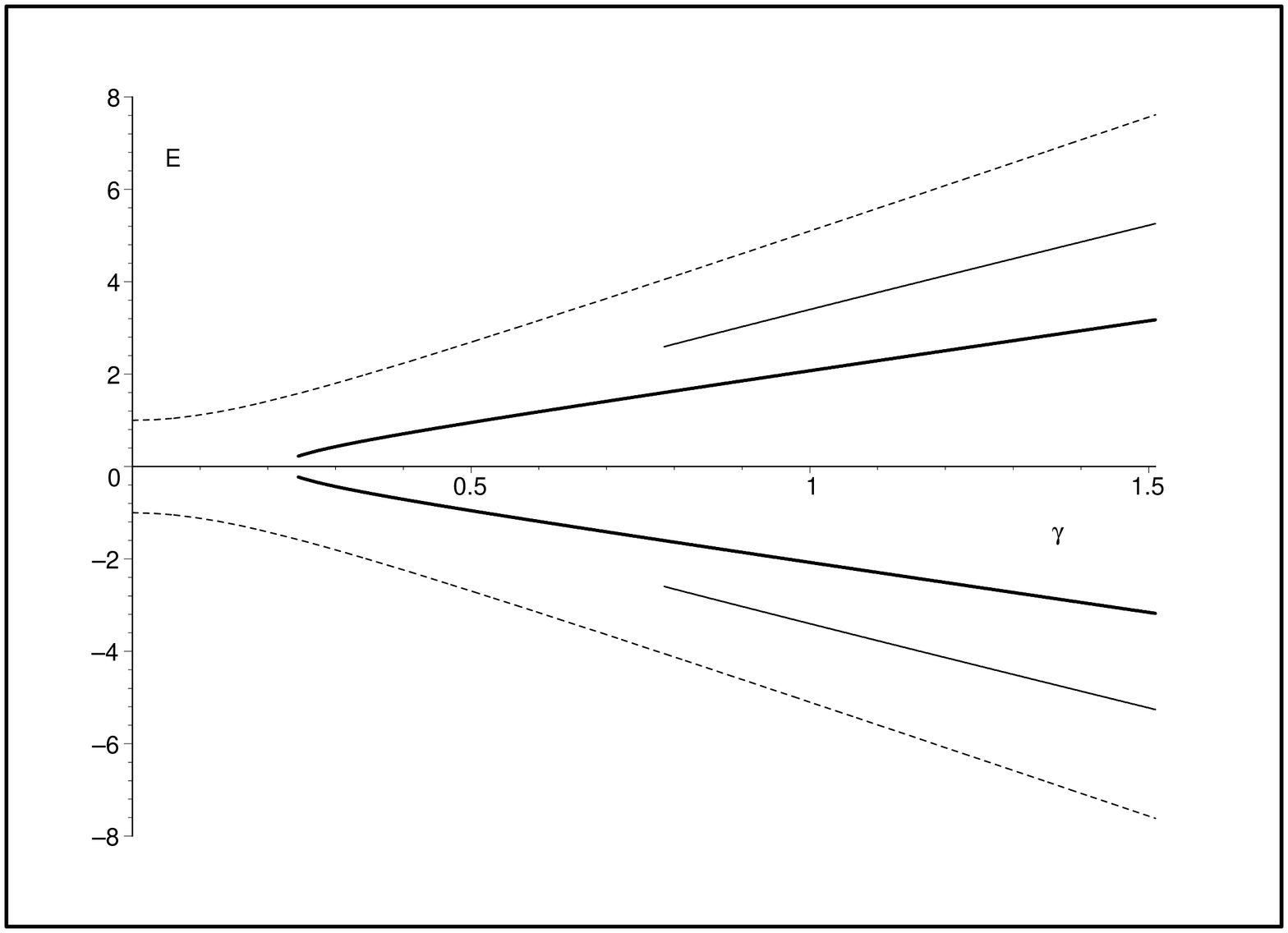}
\end{center}
\par
\vspace*{-0.1cm}
\caption{The same as in Fig. 1 as a function of $\protect\gamma $ for $%
\protect\xi =0$.}
\label{Fig3}
\end{figure}

\noindent These solutions present an intricate dependence on the parameters
of the potential and per se do not tell the whole story because the KG
energies have also to be in tune with the restrictions imposed by (\ref{13}%
). Notice that the KG energies depend only on the absolute values of the
skew parameter and of the coupling constant, and that the energy levels are
symmetric about $E=0$ for $\xi =0$ (the case of a pure scalar coupling), as
expected. The same happens for massless particles, independently of $\xi $.
This last symmetry is due to the fact that the effective eigenenergies of
the Rosen-Morse potential as well as the maximum quantum number depend on
the absolute value of $V_{2}$ and for $m=0$ this dependence transmutes into
the absolute value of $E$. Anyhow, the energy levels corresponding to the
bounded solutions are into the range $|E|<m_{\mathtt{eff}}c^{2}$. Thus, one
can see that the scalar coupling enlarges the bound-state gap and that the
particle (antiparticle) energy levels enter from the upper (lower) continuum
into the spectral gap. It is also clear that the capacity of the kink-like
potential to hold bounded solutions increases as $|g|$ and $|\gamma |$
increase. The KG energies are plotted in Figs. \ref{Fig1}, \ref{Fig2} and %
\ref{Fig3} as a function of $\xi $, $g$ and $\gamma $, respectively. The
parameters were chosen for furnishing two bounded solutions, at the most.
These figures show that the energy levels are into the range $|E|<m_{\mathtt{%
eff}}c^{2}$ and that, in general, both particle and antiparticle levels are
members of the spectrum. They also show that neither there is crossing of
levels nor the energy levels for particles (antiparticles) join the negative
(positive) continuum. These facts imply that there is no channel of
spontaneous particle-antiparticle creation so that the single-particle
interpretation of the KG equation is ensured. Fig. \ref{Fig1} shows that
both particle and antiparticle levels show their face for a pure scalar
coupling ($\xi =0$) and that they are symmetrically disposed about $E=0$.
Nevertheless, this symmetry is broken for $\xi \neq 0$ and there can be a
region where the number of particle levels is different from the
antiparticle levels. Furthermore, as $\xi \rightarrow \pm 1$ the energy
levels tend to disappear one after other. Figs. \ref{Fig2} and \ref{Fig3}
show that there are minimum values for both $|g|$ and $|\gamma |$ for the
formation of a bound-state solution.

The normalized KG wave function can be written as (see Ref. \cite{nie}):%
\begin{equation}
\phi _{n}=N_{n}\left( 1-z\right) ^{\left( a_{n}-b_{n}\right) /2}\left(
1+z\right) ^{\left( a_{n}+b_{n}\right) /2}P_{n}^{\left(
a_{n}-b_{n},a_{n}+b_{n}\right) }\left( z\right)  \label{17}
\end{equation}%
where $z=\mathrm{tanh}\,\gamma x$ and $P_{n}^{\left( \alpha ,\beta \right) }$
is the Jacobi polynomial, a polynomial of degree $n$. Furthermore,%
\begin{equation}
b_{n}=\frac{m_{\mathtt{eff}}V_{1}}{\hbar ^{2}\gamma ^{2}\left( s-n\right) }=%
\frac{g^{2}}{2a_{n}}\cos ^{2}\left( \frac{\pi \xi }{2}\right)  \label{18}
\end{equation}%
and the normalization constant is given by%
\begin{equation}
N_{n}=2^{-a_{n}}\sqrt{|\gamma |\,\frac{a_{n}^{2}-b_{n}^{2}}{a_{n}}\,\frac{%
\Gamma \left( 2a_{n}+n+1\right) \Gamma \left( n+1\right) }{\Gamma \left(
a_{n}-b_{n}+n+1\right) \Gamma \left( a_{n}+b_{n}+n+1\right) }}  \label{19}
\end{equation}

\begin{figure}[th]
\begin{center}
\includegraphics[width=10cm, angle=270]{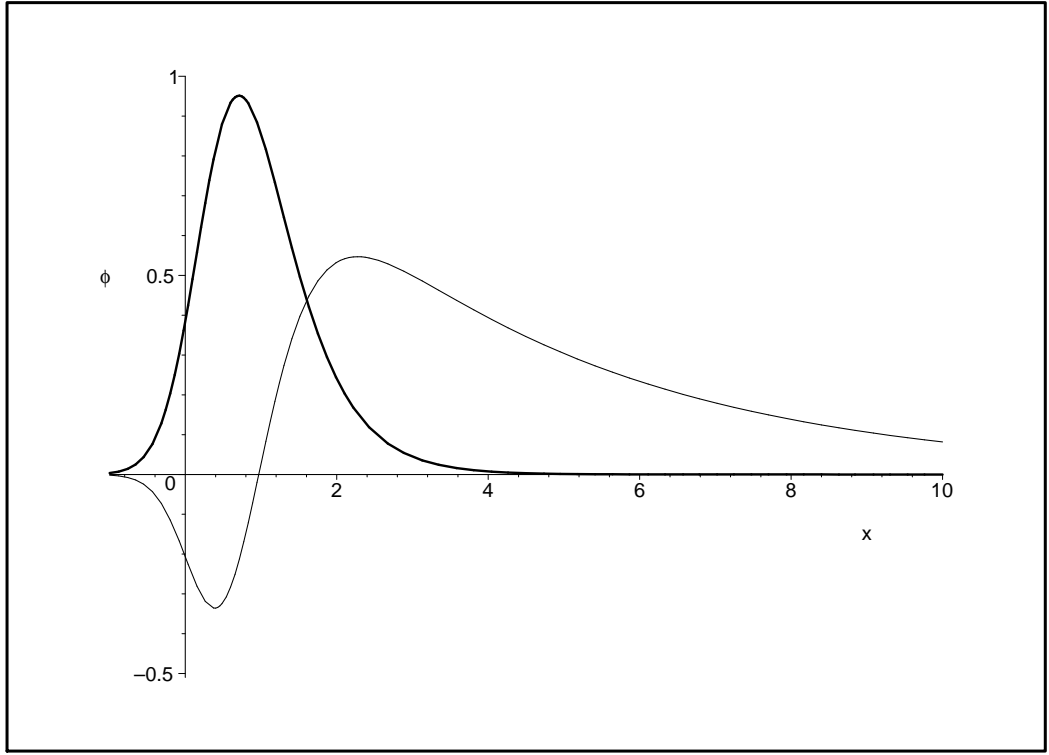}
\end{center}
\par
\vspace*{-0.1cm}
\caption{The KG wave functions as a function of the $x$ for $\protect\xi =0$%
. The thick line stands for $n=0$ and the thin line for $n=1$ ($\hbar =c=m=%
\protect\gamma =1$ and $g=5$). }
\label{Fig4}
\end{figure}

\noindent Although it may seem strange, the KG wave functions are
characterized just by $n$ so that when both particle and antiparticle levels
are present in the spectrum a pair of bound-state solutions can share the
same wave function. Indeed, this is not a riddle because the KG equation as
expressed by Eq. (\ref{1}) can not be seen as an eigenvalue problem in
general. However, $\phi $ is an eigenfunction of the operator $H_{\mathtt{eff%
}}$ defined in (\ref{5}) with $E_{\mathtt{eff}}$ as the corresponding
eigenvalue. Therefore, a pair of KG energy levels corresponding to particle
and antiparticle levels can share the same wave function if they possess the
same effective eigenenergy, i.e., the same quantum number. It is noteworthy
that the width of the position probability density, $\ |\phi |^{2}$,
decreases as $|\gamma |$ or $|g|$ increases. As such it promises that the
uncertainty in the position can shrink without limit. It seems that the
uncertainty principle dies away provided such a principle implies that it is
impossible to localize a particle into a region of space less than half of
its Compton wavelength (see, for example, \cite{str}). This apparent
contradiction can be remedied by recurring to the concept of effective
Compton wavelength defined as $\lambda _{\mathtt{eff}}=\hbar /(m_{\mathtt{eff%
}}c)$. Hence, the minimum uncertainty consonant with the uncertainty
principle is given by $\lambda _{\mathtt{eff}}/2$. It means that the
localization of a particle under the influence of the kink-like potential
does not require any minimum value in order to ensure the single-particle
interpretation of the KG equation, even if the trapped particle is massless.
As $|\gamma |$ or $|g|$ increases the binding potential contributes to
increase the effective mass of the particle in such a way that there is no
energy available to produce particle-antiparticle pairs. The KG wave
functions are displayed in Fig. \ref{Fig4}. With the chosen parameters ($%
\hbar =c=m=\gamma =1$, $g=5$ and $\xi =0$), a numerical calculation of the
uncertainty in the position of the particle for the ground-state solution
furnishes $0.475$ whereas $\lambda _{\mathtt{eff}}=0.196$.

\section{Conclusions}

For short, the bound-state solutions of the KG equation for particles
embedded in a mixed vector-scalar potential with the kink-like configuration
of the $\phi ^{4}$ model has been solved analytically. A half-half mixing
produces no bound-state solution and in fact bounded solutions only can
occur if the scalar coupling is stronger than the vector coupling. Although
the Schr\"{o}dinger equation for this sort of potential does not present any
bounded solution at all, the KG equation can present a rich spectrum which
might be useful for a better understanding of the localization of spinless
particles in the regime of strong coupling.

\vspace{3cm}

\noindent{\textbf{Acknowledgments} }

This work was supported in part by means of funds provided by CNPq and
FAPESP.

\newpage

\end{document}